\documentclass[aps,prl,twocolumn,preprintnumbers,
superscriptaddress,floatfix]{revtex4}

\usepackage{amssymb,multirow,amsmath}
\usepackage{graphicx,color}

\begin{document}

\newcommand{\Tr}{\mbox{Tr\,}}
\newcommand{\beq}{\begin{equation}}
\newcommand{\eeq}{\end{equation}}
\newcommand{\bea}{\begin{eqnarray}}
\newcommand{\eea}{\end{eqnarray}}
\renewcommand{\Re}{\mbox{Re}\,}
\renewcommand{\Im}{\mbox{Im}\,}

\title{Holographic Graphene in a Cavity}

\author{Nick Evans, Peter Jones}
\affiliation{STAG Research Centre, School of Physics and Astronomy, University of
Southampton, Southampton, SO17 1BJ, UK }
\email{evans@phys.soton.ac.uk}

\begin{abstract}
\noindent  The effective strength of EM interactions can be controlled by confining the fields to a cavity and these effects might be used to push graphene into a strongly coupled regime. We study the similar D3/probe D5 system on a compact space and discuss the gravity dual for a cavity between two mirrors. We show that the introduction of a conformal symmetry breaking length scale introduces a mass gap on a single D5 sheet. Bilayer configurations display exciton condensation between the sheets. There is a first order phase transition away from the exciton condensate if a strong enough magnetic field is applied. We finally map out the phase structure of these systems in a cavity with the presence of mirror reflections of the probes - a mass gap may form through exciton condensation with the mirror image.  
\end{abstract}

\maketitle

The low energy description of graphene \cite{Novoselov:2005kj} is given by a theory of 2+1d massless fermions interacting through classically conformal 3+1d electromagnetism. Holographic descriptions of ${\cal N}=4$ super Yang-Mills theory \cite{Maldacena:1997re,Witten:1998qj} with defect D5 probe branes \cite{Karch:2000gx,DeWolfe:2001pq,Erdmenger:2002ex,Jensen:2010ga, Evans:2010hi}
(and other related systems \cite{Skenderis:2002vf,Omid:2012vy,Grignani:2012jh,Filev:2013vka,Filev:2014bna,Filev:2014mwa}) provide a calculable system with similar gross properties - fermions on the defect interacting by higher dimension conformal gauge fields. The holographic description is only valid in the regime where the gauge fields are strongly coupled (formally the large $N$ limit of the non-abelian gauge theory).  It has been argued that the graphene system may be close to strong coupling since the effective speed of light of the fermionic theory is much less than the vacuum value \cite{Semenoff:2011jf}. Even so graphene may be in a different universality class from the holographic systems, lying closer to perturbative expectations. One way to drive graphene's interactions to stronger coupling is to place the theory in a cavity, for example placing a sheet between two mirrors. The separation of the mirrors enters the effective 2+1d electromagnetic coupling as $g_{2}^2 \sim g_3^2/L$ (see also discussion of the Purcell effect in the condensed matter literature \cite{Purcell}) and can be used to control the coupling strength. It is therefore possible that graphene could be forced into the strongly coupled regime experimentally. Holographic models may then provide useful guidance as to the expected phenomena in real world systems (although the holographic theories typically contain remnants of super-partners of the fields involved so no predictions are likely to be quantitatively correct).

Motivated by this idea, here we will study the holographic D3/probe D5 system in a compact space and in a cavity. The simplest example is to study the ${\cal N}=4$ theory in a space with one compact dimension, introducing a scale $\Delta z$. The gravity dual is an AdS-soliton configuration \cite{Horowitz:1998ha}. We place a single D5 probe in the geometry and show that a mass gap is generated by the probe brane bending in the holographic description (the system is a simple lower dimension extrapolation of the well explored D4/D6 system \cite{Kruczenski:2003uq}). This is a clean example of dynamical mass gap generation using AdS/CFT, similar to the mass gap generated by an external magnetic field \cite{Filev:2007gb}. Previously a mass gap has also been shown to develop in a system of two D5 probes in AdS$_5$ representing spatially separated defects \cite{Omid:2012vy,Karch:2005ms,Davis:2011am,Grignani:2012qz} - here the condensation occurs due to the D5 and anti-D5 branes joining in the interior of AdS and represents ``exciton'' condensation between the fermions on one defect with those on the other. In \cite{Evans:2013jma} the phase transition between that phase and the phase in a magnetic field where condensation occurs on each brane alone was investigated. Here the conformal symmetry breaking of the IR length scale is not sufficient to generate a transition but a similar transition does occur again when a magnetic field of sufficient strength is applied in addition. 

Using the ${\cal N}=4$ system to model EM interactions between mirrors is harder since it is unclear whether any true model would include such a configuration that describes both the ${\cal N}=4$ vacuum and mirrors. The AdS-soliton configuration again appears the appropriate way to introduce the IR length scale. In \cite{Fujita:2011fp} a proposal was made for AdS-duals for ${\cal N}=4$ SYM with boundaries. Amongst these proposals is one for a strip of the gauge theory between two boundaries of constant tension. The dual geometry is the AdS-soliton but with a cut off in the AdS space corresponding to the position of the boundaries (the tension of the boundary is matched to that of the plasma within). We study probe D5 branes in this system  but fail to find a regular description of a single defect since the D5 brane solutions hit the interior cut-off. Most likely this shows that the system with a tensionful wall at the edges of the strip is not a physical system that could be generated in a complete theory. However, the discussion shows that the behaviour of the gauge fields in a cavity would likely be very similar to that on the compact dimension. We therefore use the AdS-soliton to describe the vacuum state of the gauge fields and place probe D5 branes and their mirror image partners into the space. A new phase is identified in which the probe has a mass gap as a result of exciton condensation with its mirror image if the sheet comes within a quarter of the separation of the mirrors to either mirror. The complicated phase diagram for bilayer configurations is also computed in this case.

\section{Holographic ${\cal N}=4$ SYM}

We will loosely represent QED interactions by the large N dynamics of  
${\cal N}=4$ super Yang Mills theory on the surface of a stack of
D3 branes. The theory in a flat 3+1d space is described at zero temperature by AdS$_5\times
S^5$ \cite{Maldacena:1997re,Witten:1998qj}
\beq\label{ads4}
\begin{split}
ds^2  = &  {(\rho^2+L^2) \over R^2} (dx_{2+1}^2 + dz^2)\\
&+ {R^2 \over (\rho^2+L^2)} (d\rho^2 + \rho^2 d \Omega_2^2
+ dL^2 + L^2 d \tilde{\Omega}_2^2) 
\end{split}
\eeq
where we have written the geometry to display the
directions the D3 lie in ($x_{2+1},z$). A 2+1d defect with an ${\cal N}=2$ chiral multiplet on its surface can be introduced by embedding a probe \cite{Karch:2002sh} D5 \cite{Karch:2000gx,DeWolfe:2001pq,Erdmenger:2002ex,Jensen:2010ga, Evans:2010hi}
on ($x_{2+1}$, $\rho$ and $\Omega_2$) with the transverse directions $L$
and $\tilde{\Omega}_2$, plus the 3 direction that we call $z$. $R$ is the AdS radius.

\section{${\cal N}=4$ SYM on a Compact Space}

${\cal N}=4$ SYM on a space that is compact in the $z$ direction is described by the AdS-soliton \cite{Horowitz:1998ha}
\beq \label{soliton} ds^2 = {R^2 \over r^2} h^{-1}(r) dr^2 + {r^2 \over R^2} \left( dx_{2+1}^2 + h(r) dz^2 \right)  + d\Omega_5^2
\eeq
with
\beq h(r) = 1 - \left( {r_0 \over r}\right)^4 \eeq
The circumference of the space can be found by looking in the $r-z$ plane near the horizon at $r_0$ - writing $r= r_0 + \tilde{r}$ and making the transformations $\tilde{r} =  r_0 \sigma^2 / R^2$ and $\alpha = 2 r_0 z /R^2$ gives a canonical two plane metric. We impose for regularity that $\alpha$ has range $2 \pi$. Hence we learn the circumference of the $z$ direction is $R^2 \pi / r_0$.

To embed a probe-D5 brane it's convenient to make the change of coordinates
\beq w = \left( r^2 + (r^4 - r_0^4)^{1/2} \right)^{1/2} \eeq
The metric becomes
\beq ds^2 = {w^2 \over R^2} \left( g_x d x_{2+1}^2 + g_z dz^2 \right) + {R^2 \over w^2} (dw^2 + w^2 d\Omega_5^2)\eeq
where 
\beq g_x = \left(   w^4 + r_0^4 \over 2    w^4 \right)  \eeq
\beq g_z = {(w^4 - r_0^4)^2 \over  2 w^4 ( w^4 + r_0^4) } \eeq
We can now split the transverse 6-plane as before
\beq\label{ads4}
\begin{split}
ds^2  = &  {(\rho^2+L^2) \over R^2} (g_x dx_{2+1}^2 + g_z dz^2)\\
&+ {R^2 \over (\rho^2+L^2)} (d\rho^2 + \rho^2 d \Omega_2^2
+ dL^2 + L^2 d \tilde{\Omega}_2^2) 
\end{split}
\eeq

\subsection{A Single Graphene Sheet}

We will introduce quenched matter via a probe D5 brane. The matter content is a single
Dirac fermion plus scalar super partners (that will become massive in the presence of any supersymmetry breaking)
restricted to the $x_{0-2}$ directions. The 
underlying brane configuration is as follows:
\begin{center}\begin{tabular}{ccccccccccc}
& 0 & 1 & 2 & 3 & 4 & 5 & 6 & 7 & 8 & 9  \\
D3 & - & - & - & $\|$ & $\bullet$ & $\bullet$ & $\bullet$ &
$\bullet$ & $\bullet$ & $\bullet$  \\
D5 & - & - & - & $\bullet$ & - & - & - & $\bullet$ & $\bullet$ &
$\bullet$
\end{tabular}\end{center}

We expect the vacuum configuration for a single D5 probe to be described by a profile $L(\rho)$ at fixed $z$. The action for the D5 is just it's world volume 
\beq \label{action} 
\begin{split}
S &\sim -T \int d^6\xi e^\phi \sqrt{- {\rm det} G}  \\ 
  &\sim - \int d\rho~ \left( 1 + {1 \over (\rho^2 + L^2)^2}  \right)^{3/2}\rho^2 \sqrt{1 +
L^{'2} }  
\end{split} 
\eeq 
where $T$ is the tension, $\phi$ the dilaton (which is constant here as in pure AdS) and we have dropped angular
factors on the two-sphere which are a constant multiplicative factor for all the solutions we compute.  Here we have rescaled each of $L$ and $\rho$  by a factor of $r_0$.

\begin{figure}[]
\centering
\includegraphics[width=6.5cm]{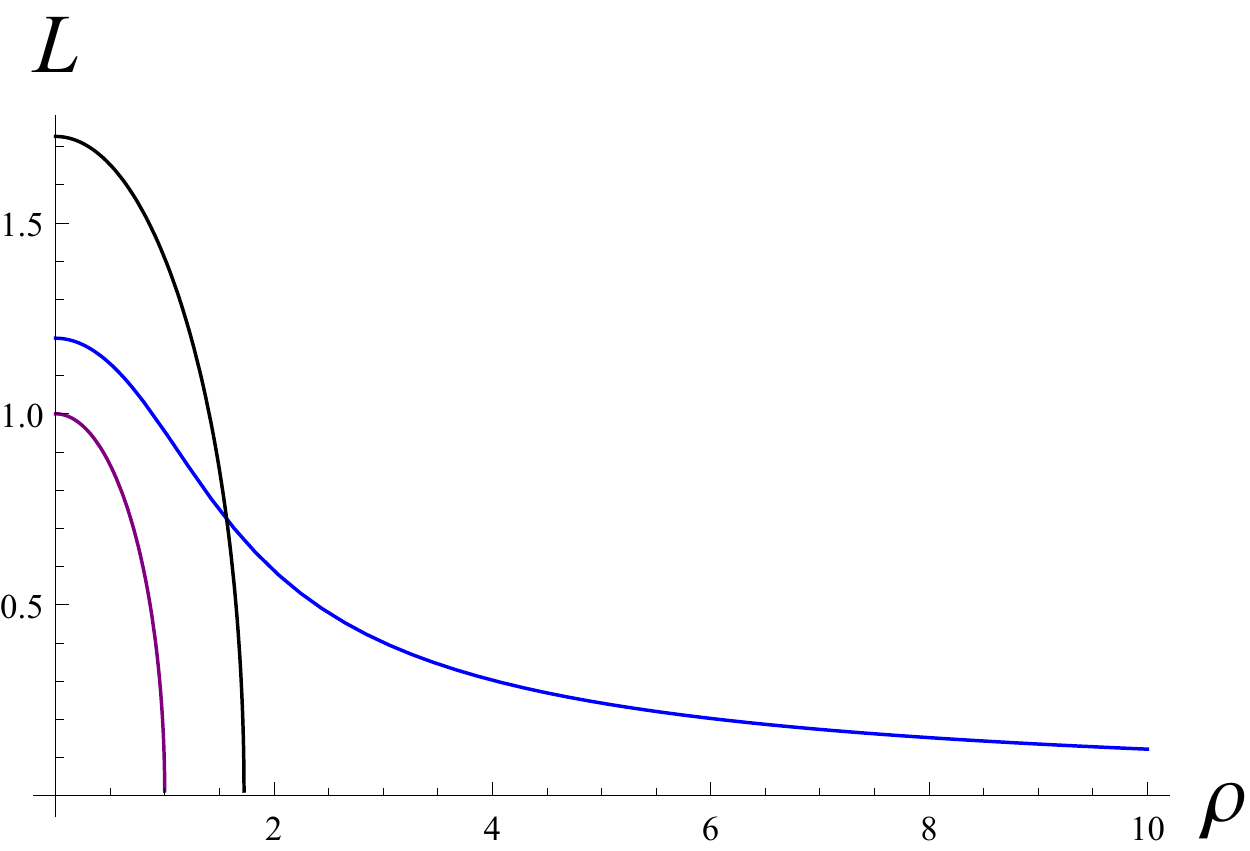} 
\caption{The embedding of a single D5 brane for a massless quark in the AdS-soliton background is shown for $r_0=1$. The dual of the compact space therefore closes off on the smaller circle shown - note the D5 brane embedding is regular avoiding this line. We also show the solution of (\ref{ta}) with the tension chosen so that the interval between the walls has radius $R^2 \pi /r_0$ - the D5 embedding hits this cut off on the space. }
\label{justL}
\end{figure}

One can solve for the numerical embedding $L(\rho)$ by shooting from $\rho=0$ subject to the boundary condition $L'(0)=0$. The asymptotic solutions fall off as $L(\rho \sim \infty) = m + c/\rho+...$ where $m$ is proportional to the quark mass and $c$ to the quark condensate (strictly it is only the quark condensate in the massless limit \cite{Skenderis:2002vf}). To find the massless embedding one shoots to find the solution with $L(\infty)=0$. The solution is shown in Fig 1 
(the position in the $\rho-L$ plane at which the geometry closes (i.e. $\omega=1$) is also shown so one sees that the embedding never enters this region). Numerical computation of $-S$ evaluated on a solution gives the vacuum energy since the configuration is static. This energy suffers from an IR divergence which goes as $\Lambda^3$ for large cutoff $\Lambda$, as can be seen from equation (9) since $L'(\rho) \rightarrow 0$ as $\rho \rightarrow \infty$ on our solution. The difference in energy between any two solutions is finite however; a simple way to regularize the solution is to subtract $\int_0^\infty \rho^2 d \rho$. 

The curved solution in Fig 1 is already an interesting result. It breaks the SO(3) symmetry of the $L=0$ embedding to 
SO(2) and the non-zero value of $L(0)$ can be interpreted as a dynamically generated IR quark mass. This response is familiar from other cases in the literature describing the dynamical generation of a mass gap \cite{Babington:2003vm,Kruczenski:2003uq,Filev:2007gb}. This graphene like configuration on a compact space provides an example of a completely controlled AdS/CFT computation of a dynamically generated mass gap (the case with magnetic field is the only other known case for a single probe D5 brane \cite{Jensen:2010ga, Evans:2010hi}).

\subsection{Bilayer Configurations}

Another example of a holographically computable dynamical mass gap is provided by a bilayer configuration of probe D5 branes in AdS \cite{Skenderis:2002vf,Omid:2012vy,Grignani:2012jh,Filev:2013vka,Filev:2014bna,Filev:2014mwa, Evans:2013jma}. A joined D5/anti-D5 brane U-shaped configuration, analogous to the Wilson loop configuration of an interacting quark and anti-quark
\cite{Maldacena:1998im,Rey:1998bq}, describes this  ``exciton'' condensation between the fermions on the two defects. The separation of the defects provides the conformal symmetry breaking scale.  It is interesting to explore this configuration in the compact space therefore. 

We will allow for an embedding of the probe D5 brane in the $z$ direction. Allowing $z$ to depend on $\rho$ only, the action for the case $L=0$ now becomes
\beq   \label{action2}
S \sim -  \int d\rho~ \left( 1 + {1 \over \rho^4}  \right)^{3/2}\rho^2 \sqrt{1 +
\frac{(\rho^4-1)^2}{2(\rho^4+1)}z^{'2}}  \,
\eeq 
We have again scaled $\rho$ by $r_0$ and now $z$ by a factor of $R^2/r_0$ (the circumference of the compact $z$-direction is now $\pi$ in these coordinates). 
In general there may be solutions with non-zero $z$ and $L$ simultaneously but we shall not consider such configurations
(those configurations were explored in \cite{Evans:2013jma} for the B-field case but shown to only represent local maxima of the effective potential).

Since the action is independent of $z$, there is a constant of motion $\Pi_z$ given by
\beq
\Pi_z = \frac{z' \sqrt{1+1/\rho^4}(\rho^4-1)^2}{\rho^2 \sqrt{2+z'^2(\rho^4+4/(1+\rho^4)-3)}}
\eeq

As mentioned, these solutions represent two branes joining, and extend a finite distance into the bulk of the space at which point they turn around. There thus exists some $\rho_0$ at which $z' \rightarrow \infty$. From equation (11) one then finds that
\beq \label{pi1}
\Pi_z = \frac{\sqrt{1+1/\rho_0^4}(\rho_0^4-1)^2}{\rho_0^2 \sqrt{\rho_0^4+4/(1+\rho_0^4)-3}}
\eeq
From equation (11) one also finds
\beq \label{pi2}
z' = \pm \frac{\sqrt{2}\Pi_z \rho^2}{\sqrt{(1+1/\rho^4)(\rho^4-1)^4-\Pi_z^2\rho^4(\rho^4+4/(1+\rho^4)-3)}}
\eeq
giving a first order ODE for $z(\rho)$ which one can directly integrate up numerically. One can also find the separation $\Delta z$ of the branes in a given solution by integrating equation (13) over $\rho \in [\rho_0, \infty]$. The solutions for various values of $\rho_0$ are shown in Fig 2.

\begin{figure}[]
\centering
\includegraphics[width=6.5cm]{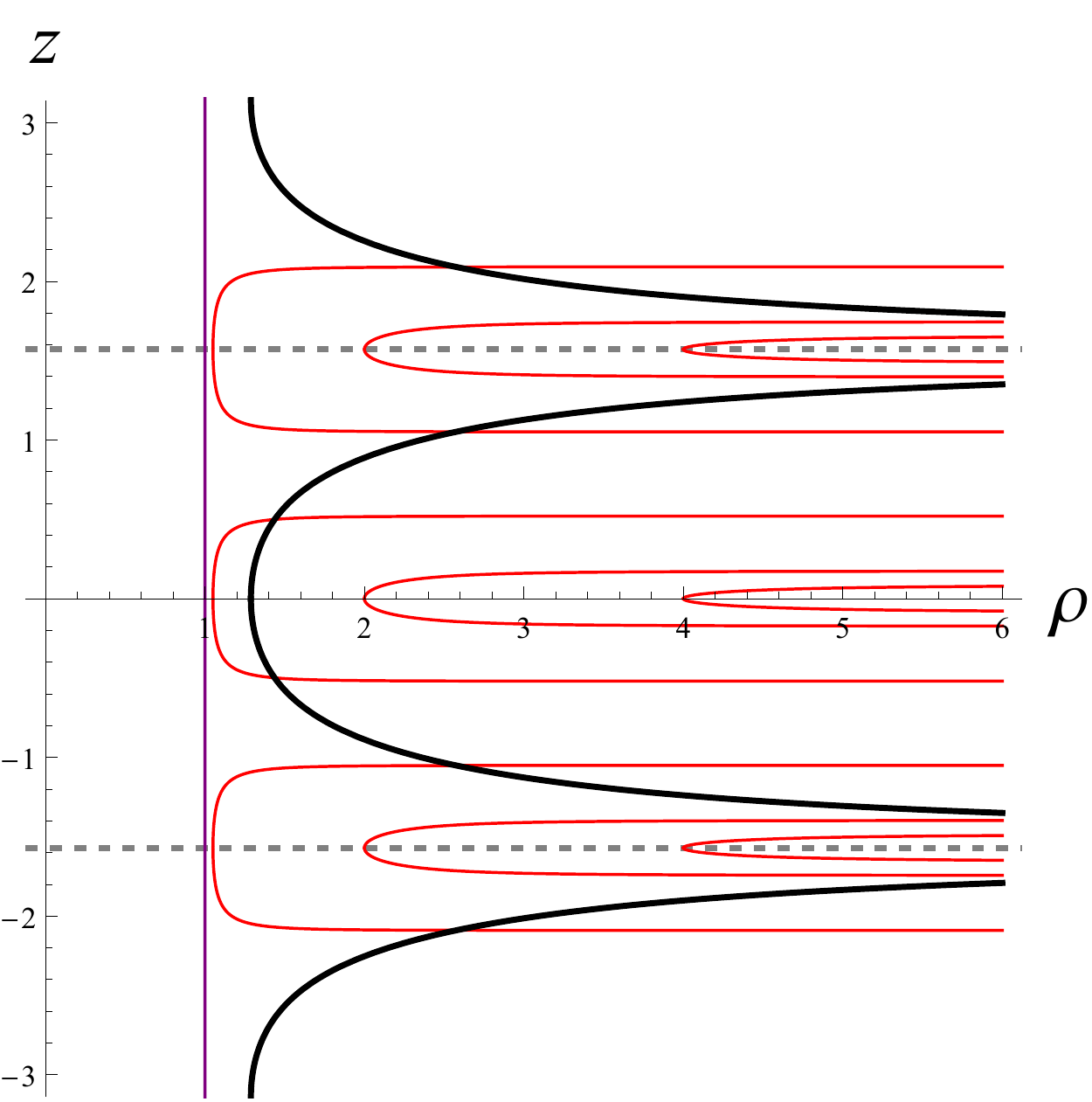} 
\caption{U-shaped D5/$\bar{D5}$ configurations in the AdS-soliton background. The space has circumference $\pi$ in the $z$-direction - we repeat the space to show configurations wrapping both ways around the circle. Note that configurations which reach down to $r_0=1$ describe defects separated by $\pi/2$ in $z$. We also plot the solution of (\ref{ta}) with the tension chosen so that the interval between the walls is $\pi$ - the probe configurations hit this boundary. If the figure is viewed as that for an interval between two mirrors then configurations corresponding to exciton condensation with the mirror partner exist if the probe lies within $\pi/4$ of the mirror.}
\label{U}
\end{figure}

\begin{figure}[]
\centering
\includegraphics[width=6.7cm]{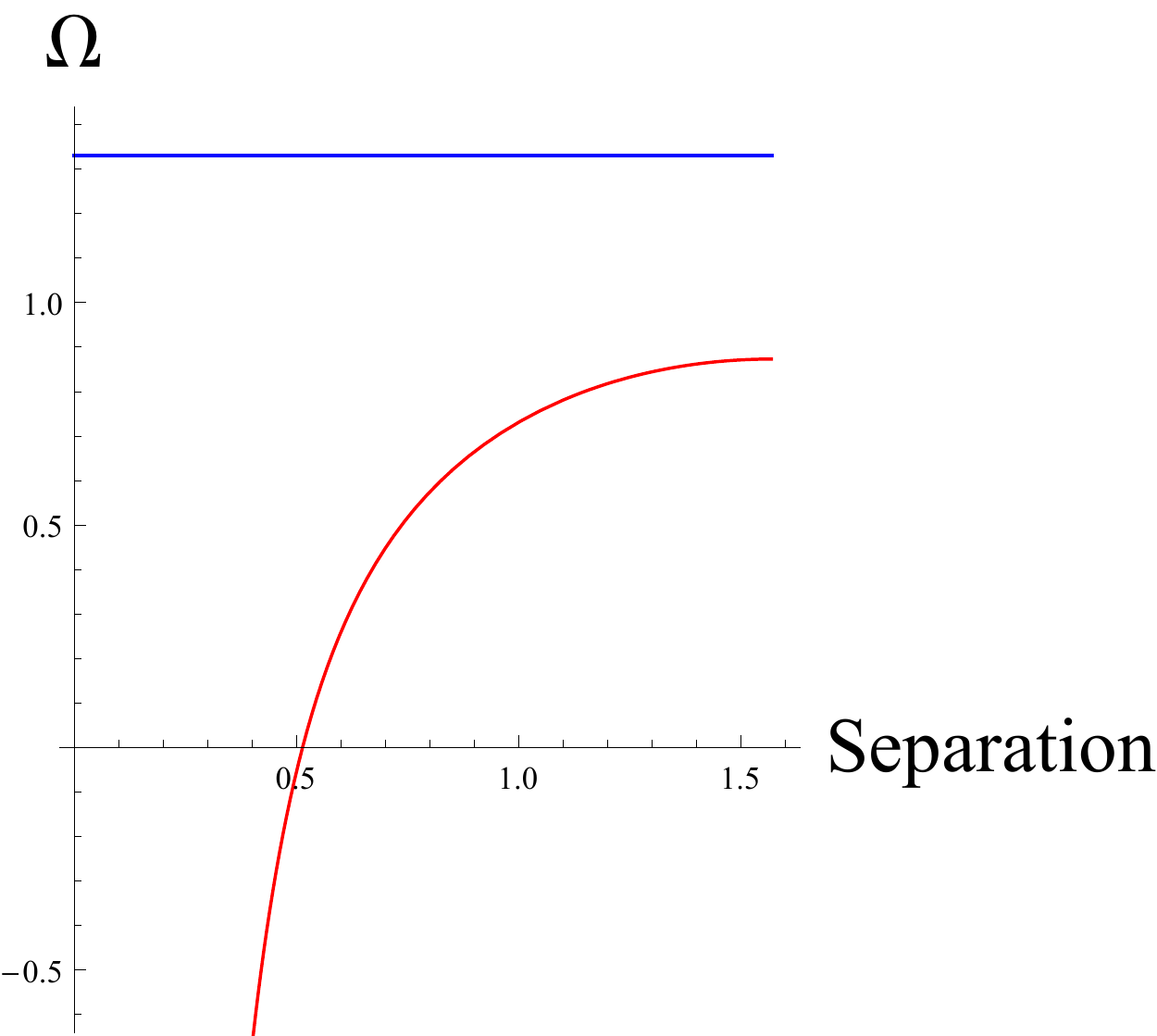} 
\caption{The regularized energy of two D5s in the configuration of Fig \ref{justL} (which is independent of $z$) and that of the U-shaped joined embedding against separation in $z$ (which cannot exceed $\pi/2$). The U shaped configurations are always preferred. }
\label{energy}
\end{figure}

An interesting feature is that there are no linked configurations with separation greater than $\pi/2$ (half the circumference of the compact direction). The maximum $\pi/2$ separation is realized precisely when the linked configurations falls into the scale $r_0$. For a compact space the dual has provided us with precisely the minimum number of configurations to describe all possible probe D5 separations - two D5 probes can not be separated by more than half the circumference of the $z$-direction. The solutions do not provide us with configurations that wind further around the compact direction though. 

For the purposes of computing the energies of these solutions, the expression given in equation (13) can be substituted directly into the action in (10). The asymptotic behaviour of the on-shell action can then easily be read off, and we again subtract $\int_0^\infty \rho^2 d \rho$  to regularize. The results for the two embeddings are given in Fig \ref{energy}. The linked embedding always has lower energy than the separated embeddings - exciton condensation between the sheets is preferred over separated sheets with condensation on the single sheet at all separations on the $z$ circle.

\subsection{Applying a Magnetic Field}

For the compact space, U-shaped probe configurations were always energetically preferred over the configuration in Fig \ref{justL} for bi-layers. 
It is known that the configuration of Fig 1 though is heavily preferred when a large background constant magnetic field is applied \cite{Filev:2007gb,Jensen:2010ga, Evans:2010hi}. Here we mean a B-field associated with the U(1) baryon number symmetry which has a dual description in terms of a gauge field in the DBI action of the brane and not a component of the ${\cal N}=4$ SYM fields.
The full DBI action with that gauge field is given by
\beq
S \sim - T \int d^6\xi e^\phi \sqrt{- {\rm det} [G+2 \pi \alpha' F]}
\eeq
For the 2+1-dimensional defect field theory in question, the only possibility for introducing an external magnetic field is via $F_{yx}=-F_{xy}=B_z \equiv B$. Following through the analysis as before, one finds that the actions retain the forms (\ref{action}) and (\ref{action2}) but with an additional factor that can be identified as an effective dilaton profile
\beq
e^{\phi}=\sqrt{1+\frac{(2 \pi \alpha')^2B^2}{g_x^2 (\rho^2+L^2)^2}}
\eeq

Equations (\ref{pi1}) and (\ref{pi2}) for the bilayer configurations now become
\beq
\Pi_z = \frac{\sqrt{1+1/\rho_0^4}(\rho_0^4-1)^2\sqrt{1+4B^2\rho_0^4/(1+\rho_0^4)^2}}{\rho_0^2 \sqrt{\rho_0^4+4/(1+\rho_0^4)-3}}
\eeq
and
 \beq
z' = \frac{\sqrt{2}\Pi_z \rho^2}{\sqrt{{\cal A} -{\cal B}}} 
\eeq
\beq \begin{array}{ccc}
{\cal A} &=& (1+1/\rho^4)(\rho^4-1)^4\left(1+\frac{4B^2\rho^4}{(1+\rho^4)^2}\right)\\
&&\\
{\cal B} &=& \Pi_z^2\rho^4(\rho^4+4/(1+\rho^4)-3)\end{array}
\eeq
where we are writing $B$ in units of $1/2\pi\alpha'$. One finds that for any value of $B$, the maximum possible separation of the branes (corresponding to $\rho_0=1$) is again given by $\pi/2$. Of course, the equation of motion for $L$ also changes accordingly, and for a given value of $B$ one again shoots numerically to find the solution with $L(\infty)=0$, corresponding to the massless embedding. 

\begin{figure}[]
\centering
\includegraphics[width=7.8cm]{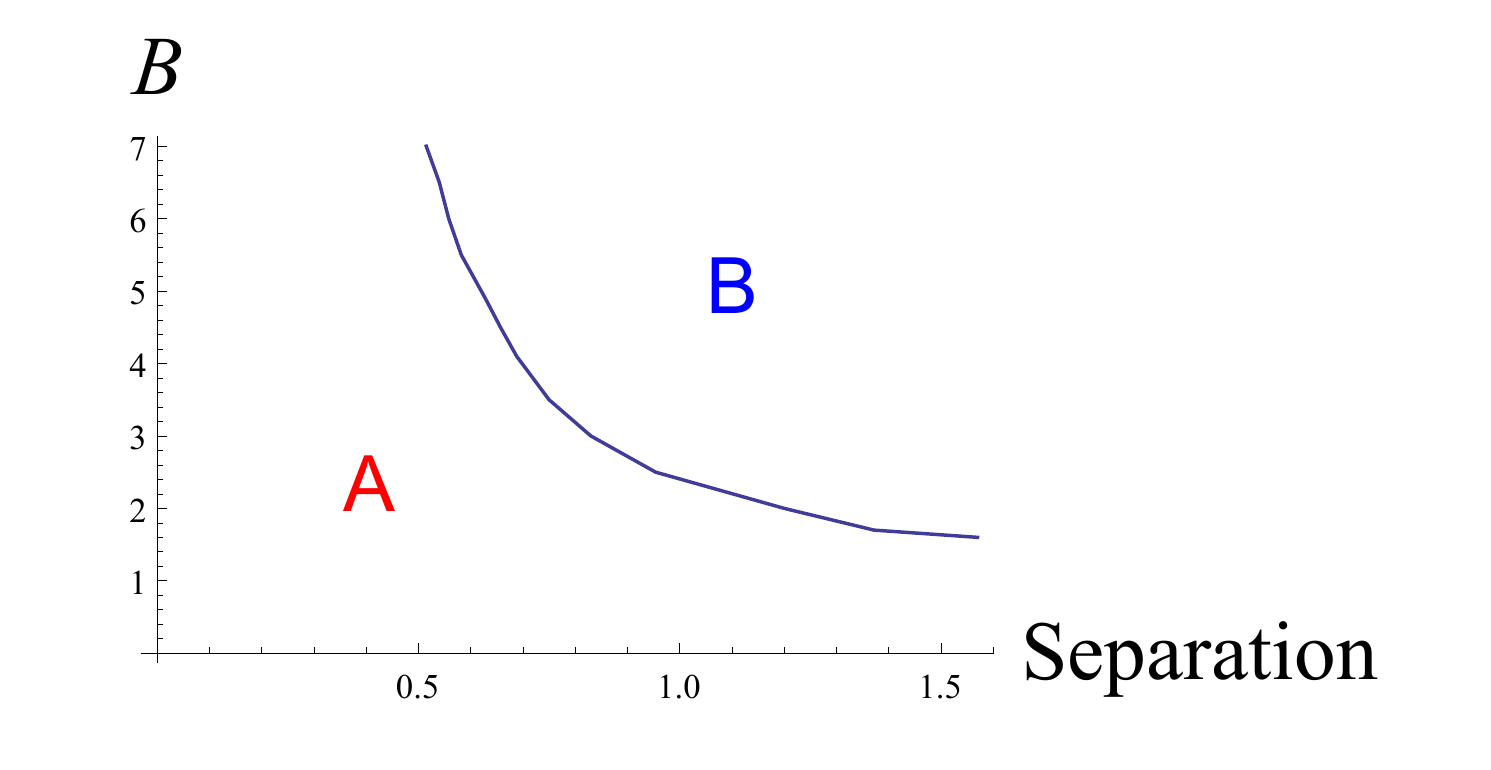} 
\caption{Phase diagram of bi-layer D5s in AdS with one compact direction and an applied B field. Phase A is U-shaped configurations. In Phase B the probes separate and take up configurations similar to Fig \ref{justL}.}
\label{Bphase}
\end{figure}

For both types of embeddings the same regularisation as before holds, as the new factor introduces no new divergences. The energies of the solutions are increased as one increases the strength of the magnetic field, but this does not happen at the same rate so  that a phase transition exists above a certain value of $B$. The phase diagram of the theory is shown in Fig \ref{Bphase} - at low B the U-shaped configurations are preferred (phase A in the figure) whilst at large B the separated brane configurations has lowest energy (phase B in the figure). The closer the probe branes the larger the B-field needed to trigger the transition.

\section{${\cal N}$=4 SYM in a cavity}

We would like to describe the background ${\cal N}$=4 SYM fields enclosed in a cavity, for example between two mirrors
placed $\Delta z$ apart. If the vacuum of the theory is locally determined then one would expect the description of the vacuum state to match that of the theory on a compact space since one can consider the space to simply repeat every $\Delta z$ period. The soliton geometry of (\ref{soliton}) seems the good candidate. Whether this is true or not depends on the boundary where there can potentially be boundary terms associated with the ``mirror". For example in  \cite{Fujita:2011fp} the authors proposed treating the surface as a boundary of constant tension ${\cal T}$. The bulk plus boundary action is then
\begin{equation}\begin{array}{ccl}
I &= &{1 \over 16 \pi G_N}\int_{\rm bulk} \sqrt{-g} (R - 2 \Lambda)\\ && \\ &&+ {1 \over 8 \pi G_N} \int_{\rm bound} \sqrt{-h} (K-{\cal T}) \end{array}
\end{equation}
where $K_{ab}$ is the extrinsic curvature. The boundary condition 
\begin{equation}
K_{ab} = (K - {\cal T}) h_{ab} \end{equation}
results. For the case of a strip between two such boundaries this gives the differential equation
\begin{equation} \label{ta}
z'(r) = \pm {R{\cal T} \over r^2 h(r) \sqrt{4 h(r) - R^2 {\cal T}^2}} \end{equation}
where the sign depends on which side of the strip the boundary sits. The solutions of this equation represent the radial evolution of the position where the tension of the boundary matches the pressure of the interior gauge fields. The solutions are U-shaped dipping down to a distance $r_*$ in the bulk. At the mid-point $4 h(r_*) = R^2 T^2$. One can then integrate (\ref{ta}) from $r=r_*$ to $\infty$ and require that $\Delta z$ matches that from the regularity condition of the geometry (\ref{soliton}) - this fixes the tension $T$.  

We plot the solutions of (\ref{ta}) in Fig \ref{justL} and Fig \ref{U} above. They represent a cut off on the space in this construction. We would now be interested in including probe D5 branes in the space. The Euler-Lagrange equations describing the embeddings in the new space are identical to those in the compact space and we would hope the probes to close off before hitting the new boundaries. In Fig \ref{justL} and Fig \ref{U} we see that this is not the case. It is possible that one needs to invoke new boundary conditions when the probes meet the cut-off to reflect the physics of the interaction between the fields on the probes and the barrier physics. Equally likely though is that the construction does not make complete sense. One is attempting to construct a theory with mirrors and a region of space with the vacuum of ${\cal N}$=4 SYM but such a theory may well not exist because the matter needed to construct the mirror would need to be part of the vacuum in the strip.

It is not our intention to resolve these complex issues here. Instead we will take the most naive prescription of simply using the soliton geometry as our description of the vacuum of the theory between two mirrors and place probe branes with their mirror images in that space. The embedding solutions are then simply those we have already displayed for the compact space. We hope that this will reveal the qualitative new physics correctly. 
 
Let us first take this approach for a single probe D5 brane between the mirrors. The immediate assumption is that the configuration is that shown in Fig 1 - there will be condensation of the fermions on the brane triggered by the conformal symmetry breaking scale $\Delta z$. However, there is an additional interesting possibility which is that there can be exciton style condensation with the mirror images of the probe. When the probe is at the mid-point between the mirrors it is a distance $\Delta z$ from it's reflections. There are no U-shaped configurations for probes separated by more than $\Delta z/2$ so the single configuration of Fig 1 is correct. If the probe though is moved within $\Delta z/4$ of the mirror then joined configurations exist with lower energy than the single configuration and exciton condensation with the mirror image will occur. In a sense this is a new form of mass gap formation for this system. 

\begin{figure}[]
\centering
\includegraphics[width=7.8cm]{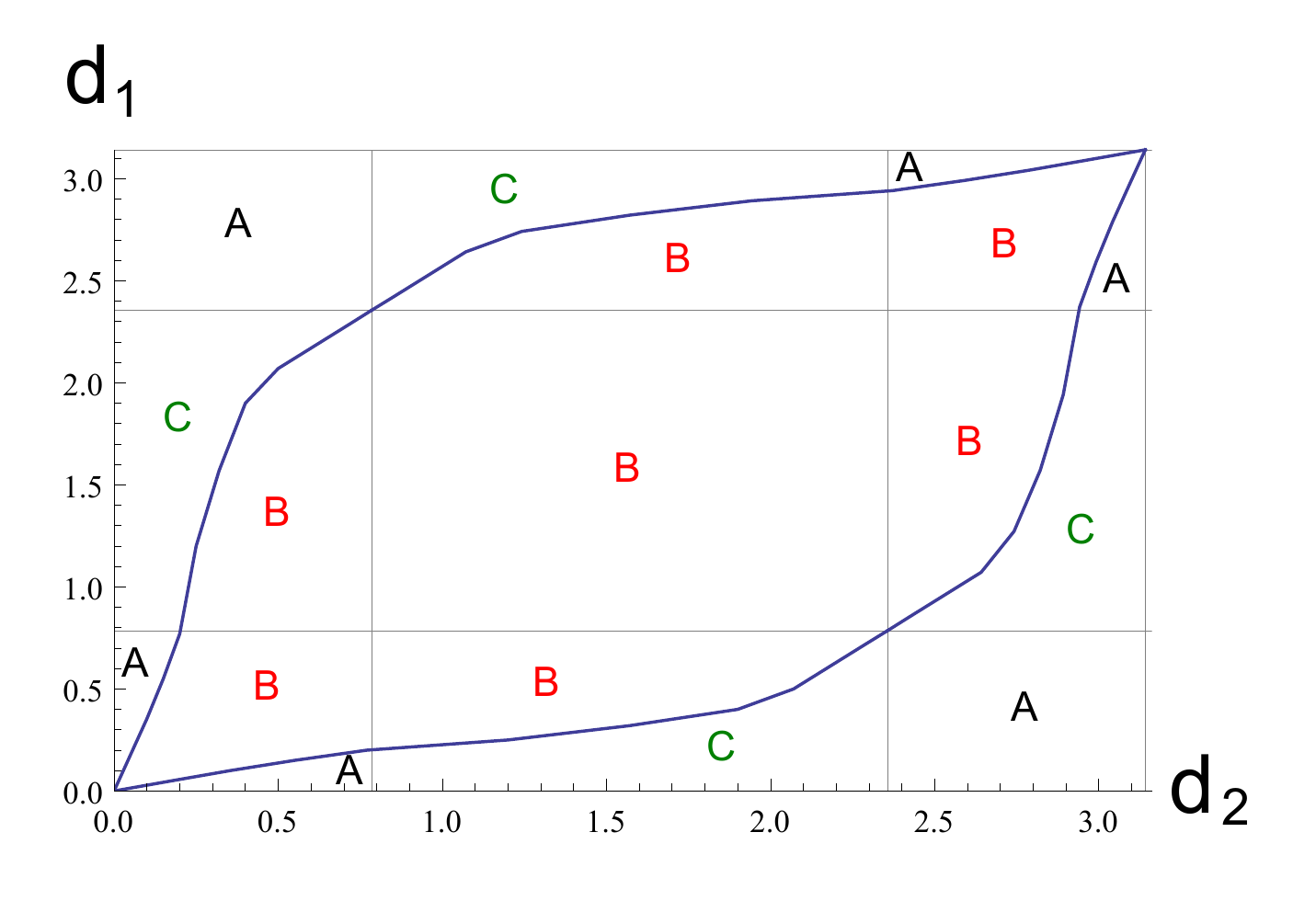} 
\caption{The phase diagram of the bilayer theory in an interval between two mirrors of separation $\pi$. $d_1$ and $d_2$ measure the distance from one mirror to the first and second defect respectively. We have marked the lines $d_{1/2}=\pi/4, 3 \pi/4$ because these are the separations within which condensation with the mirror image are possible. In phase A both D5s condense with their mirror images. In phase B the two D5s form a U-shaped configuration. In phase C the probe nearest the mirror displays exciton condensation with its mirror partner whilst the other probe takes up the lone configuration of Fig \ref{justL}. }
\label{stripbiphase}
\end{figure}

For two probe branes several configurations are possible depending on the separation of the branes from each other and from their mirror partners. Both branes can condense with their mirrors, one can condense with the mirror and one have condensation of only its own fields, or the two branes can display exciton condensation with each other. One works through all possibilities and computes the enrgetically preferred configuration. The phase diagram is shown in Fig \ref{stripbiphase}. 

\section{Discussion}

We have studied D5 probe embeddings in an AdS-soliton configuration. The geometry is dual to ${\cal N}=4$ SYM with one spatial direction of the 3+1d space compact. We have also argued that it is dual to the vacuum of the theory of ${\cal N}=4$ SYM confined to a compact region between two mirrors (although as we discussed this is ambiguous and essentially assumes the mirrors do not contribute to the form of the vacuum configuration of the ${\cal N}=4$ fields except through the introduction of a length scale). The conformal symmetry breaking scale introduced through the finite distance in $z$ in both cases generates fermion condensation and mass gap formation. For a compact space a single defect exhibits a fermion condensate on its surface. For a bilayer D5/$\bar{D5}$ configuration the energetically preferred condensation is exciton condensation between the fermions on the two sheets. If one includes mirror images of the probes in the case of the interval then an extra phase appears in which a single fermion, when close enough to the mirror, displays exciton condensation with its mirror image. The bilayer phase structure is then considerably complicated (we display the phase diagram in Fig \ref{stripbiphase}).   

The hope is that graphene sheets can be engineered into a strongly coupled phase by placing them in a cavity. The qualitative expectations from our results are that a mass gap will form in this regime. Further predictions would then be that there would be a first order phase transition to condensation with the mirror partner if the single sheet were brought close to the mirror. For bi-layer configurations we also showed that an applied magnetic field can cause a first order transition from an exciton condensation phase to a phase with separate condensation on each sheet. Potentially these sorts of features could be looked for experimentally.

\noindent {\bf Acknowledgements:} The authors would like to thank Simeone de Liberato for suggesting that graphene could be forced to strong coupling in a cavity. We thank Jegor Korovins for directing us to \cite{Fujita:2011fp}.
NE is grateful for the support of STFC.

\end{document}